\documentclass[conference]{IEEEtran}
\usepackage{cite}
\usepackage[hyphens]{url}
\usepackage{amsmath,amssymb,amsfonts}
\usepackage{algorithmic}
\usepackage{graphicx}
\usepackage{textcomp}
\usepackage{xcolor}

\def\BibTeX{{\rm B\kern-.05em{\sc i\kern-.025em b}\kern-.08em
    T\kern-.1667em\lower.7ex\hbox{E}\kern-.125emX}}
\usepackage[letterpaper]{geometry}
\begin{document}

\title{Automated Quantification of Traffic Particulate Emissions via an Image Analysis Pipeline\\
\thanks{*Equal contribution}
}

\author{
\IEEEauthorblockN{Kong Yuan Ho*}
\IEEEauthorblockA{\textit{Dept of Civil Engineering} \\
\textit{National University of Singapore}\\
Singapore, Singapore \\
barronkyho@gmail.com}
\and
\IEEEauthorblockN{Chin Seng Lim*}
\IEEEauthorblockA{\textit{Dept of Civil Engineering} \\
\textit{National University of Singapore}\\
Singapore, Singapore \\
limchinseng98@gmail.com}
\and
\IEEEauthorblockN{Mathena A Kattar}
\IEEEauthorblockA{\textit{Dept of Civil \& Environ. Engineering} \\
\textit{National University of Singapore}\\
Singapore, Singapore \\
e0325813@u.nus.edu.sg}
\and
\IEEEauthorblockN{Bharathi Boppana}
\IEEEauthorblockA{\textit{Dept of Fluid Dynamics} \\
\textit{Inst. of High Performance Computing}\\
Singapore, Singapore \\
boppanavbl@ihpc.a-star.edu.sg}
\and
\IEEEauthorblockN{Liya Yu}
\IEEEauthorblockA{\textit{Dept of Civil \& Environ. Engineering} \\
\textit{National University of Singapore}\\
Singapore, Singapore \\
liya.yu@nus.edu.sg}
\and
\IEEEauthorblockN{Chin Chun Ooi}
\IEEEauthorblockA{\textit{Dept of Fluid Dynamics} \\
\textit{Inst. of High Performance Computing}\\
\textit{Center for Frontier AI Research}\\
Singapore, Singapore \\
ooicc@ihpc.a-star.edu.sg}
}

\maketitle

\begin{abstract}
Traffic emissions are known to contribute significantly to air pollution around the world, especially in heavily urbanized cities such as Singapore. It has been previously shown that the particulate pollution along major roadways exhibit strong correlation with increased traffic during peak hours, and that reductions in traffic emissions can lead to better health outcomes. However, in many instances, obtaining proper counts of vehicular traffic remains manual and extremely laborious. This then restricts one's ability to carry out longitudinal monitoring for extended periods, for example, when trying to understand the efficacy of intervention measures such as new traffic regulations (e.g. car-pooling) or for computational modelling. Hence, in this study, we propose and implement an integrated machine learning pipeline that utilizes traffic images to obtain vehicular counts that can be easily integrated with other measurements to facilitate various studies. We verify the utility and accuracy of this pipeline on an open-source dataset of traffic images obtained for a location in Singapore and compare the obtained vehicular counts with collocated particulate measurement data obtained over a 2-week period in 2022. The roadside particulate emission is observed to correlate well with obtained vehicular counts with a correlation coefficient of 0.93, indicating that this method can indeed serve as a quick and effective correlate of particulate emissions.  
\end{abstract}


\section{Introduction}
Vehicular emissions is one of the major contributors to air pollution in urban cities such as Singapore and a potential health hazard \cite{buckeridge2002effect,quah2003economic}. To improve air quality, many governments, including the government of Singapore, have implemented regulations on emissions and fuels of vehicles to reduce the impact from motor vehicles, such as the Vehicular Emissions Scheme (VES) \cite{nea|air_quality}. Traffic management schemes such as car-pooling during peak hours and congestion-based tolls have also been implemented and studied in an attempt to reduce vehicle density on the roads \cite{shewmake2012can,daniel2000environmental}. 

Accurate study of potential particulate matter (PM) emissions from vehicular traffic depends critically on accurate estimation of their numbers, and knowledge of their average emissions. Several studies have thus been conducted to better obtain databases of potential emissions for vehicles of various types and ages, spanning different countries \cite{cai2007estimation,davis2005development,pokharel2002road}. Accurate quantification of these emissions can in turn be useful as boundary conditions for the modelling and study of potential traffic pollutant dispersion and possible interventions such as air filters and vegetation barriers \cite{boppana2019cfd,konczak2021assessment}. However, many studies currently still resort to manual counting of the vehicular traffic when attempting to quantify the traffic density for correlation with particulate levels \cite{velasco2016particles,zheng2021impacts}.

Hence, in this work, we seek to develop a fully automated image processing pipeline for quantifying the number of vehicles passing through a road segment, capitalizing on the wealth of foundational computer vision models that have been made freely available. This automated quantification can then be flexibly applied to quantification of vehicular counts for the different kinds of traffic congestion and pollutant studies described above, including in combination with other pollutant sensors. Critically, we propose the use of an open-source image segmentation and classification model to facilitate easy adoption and application by the average researcher for their own studies.

As a proof-of-concept, we assess the feasibility and effectiveness of this automated analysis pipeline by using this pipeline for the study of a set of open-source traffic images obtained for a location in Singapore. We correlate the obtained vehicular counts with particulate matter measurements at the same location as a means of validating the developed pipeline, and obtain good correlation ($>$ 0.9).

Critically, while this paper focuses on verifying the relationship between traffic density and roadside particulate concentration levels for one specific location in Singapore, we believe the automated pipeline outlined here to be easily extended to other image sources and locations.

\section{Methods}

\subsection{Traffic Image Dataset}
Instantaneous traffic emissions are generally estimated by counting the number of vehicles on the road and applying a conversion factor based on estimates of the typical emission for each vehicle class at an estimated velocity. Typically, one can obtain such traffic density estimates via a count of vehicles on the road. 

\begin{figure}[htbp]
\begin{center}
\centerline{\includegraphics[width=0.9\linewidth]{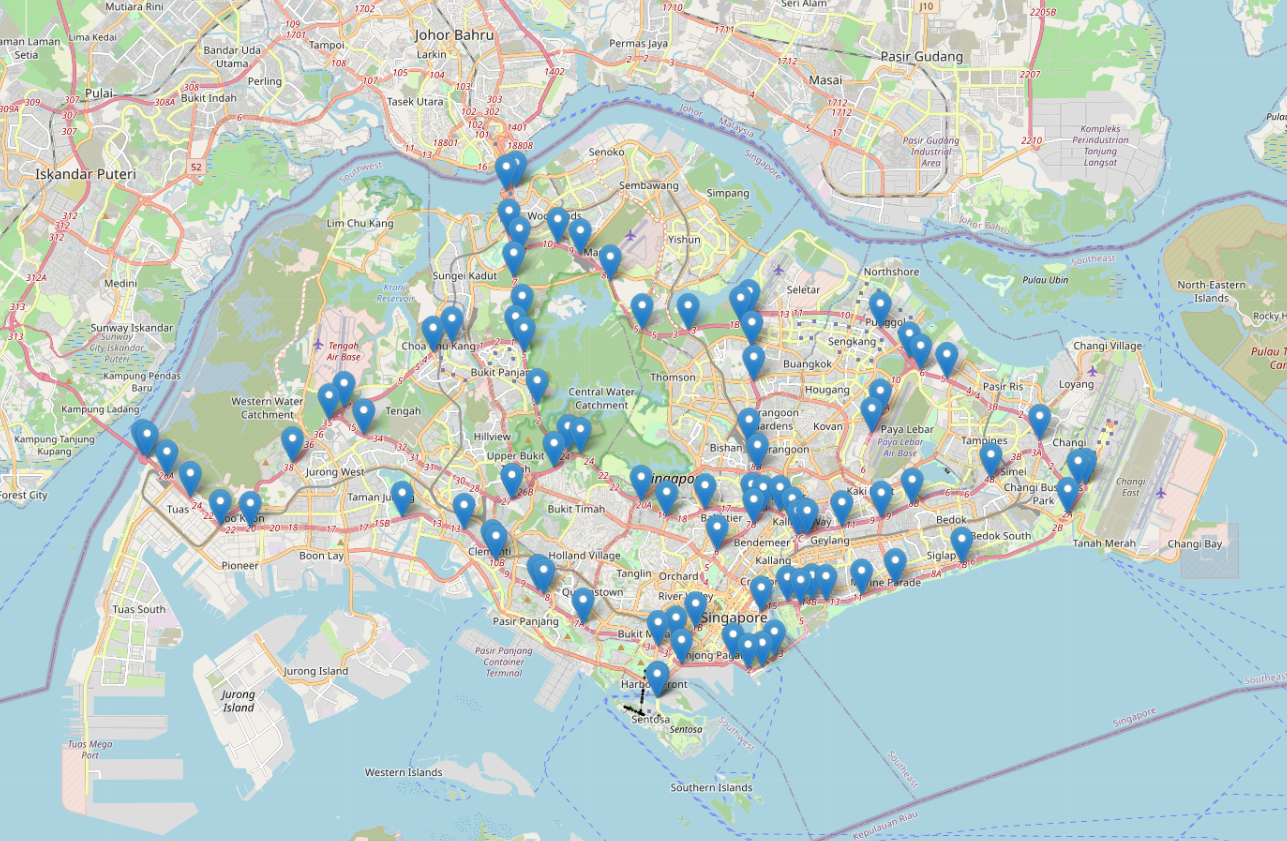}}
\caption{Illustration of open-source traffic camera locations in Singapore}
\label{Camera 1}
\end{center}
\vspace*{-7mm}
\end{figure}

In this work, we propose to use images acquired by a permanent roadside camera installation that have been made publicly available by the Government of Singapore. These images are extracted from Singapore's Land Transport Authority's network of cameras that have been made publicly available on https://data.gov.sg. Figure \ref{Camera 1} shows the locations of all the traffic cameras in this open source data repository. 

\begin{figure}[htbp]
\begin{center}
\centerline{\includegraphics[width=0.9\linewidth]{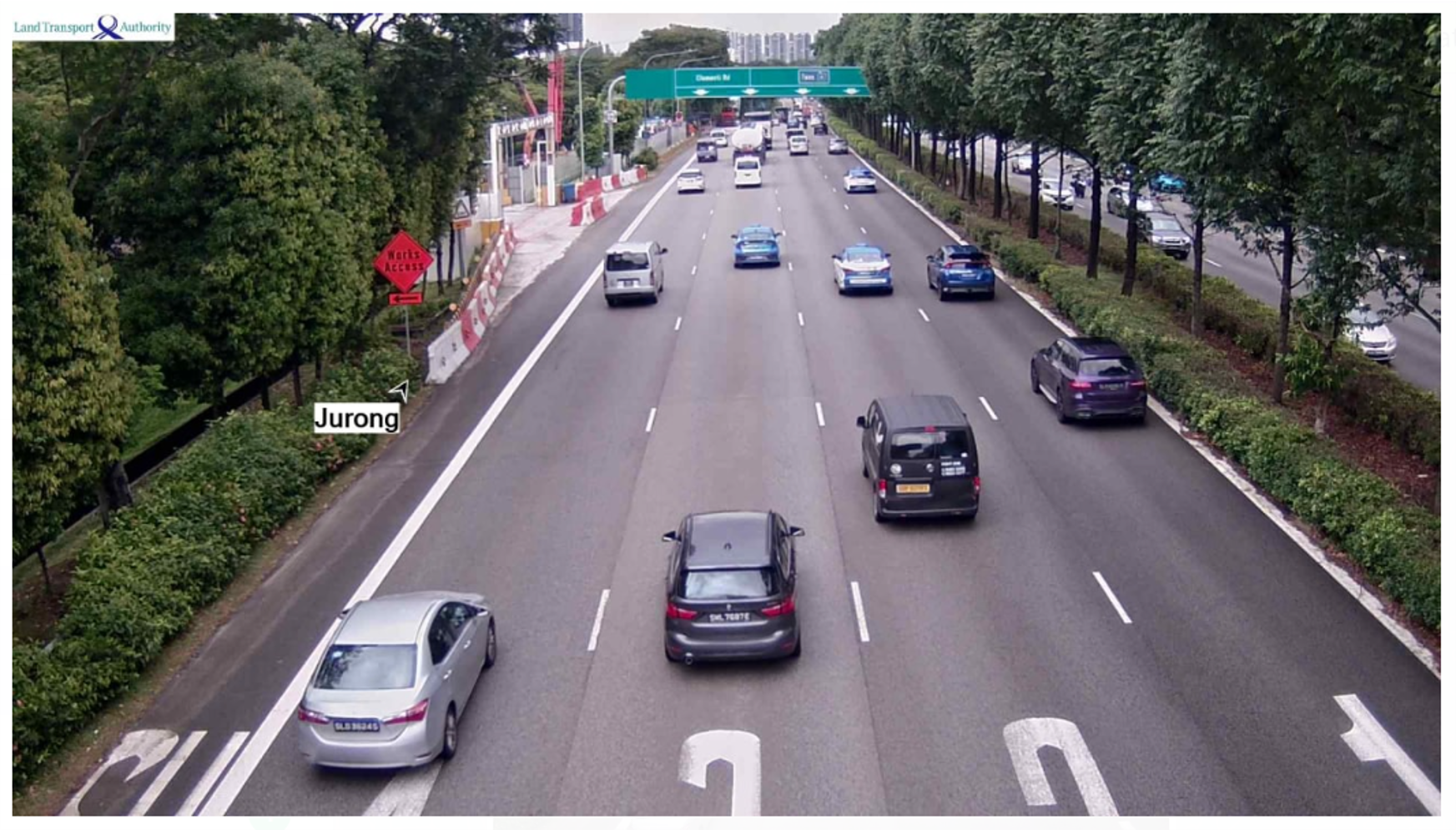}}
\caption{Sample traffic image from Singapore's Land Transport Authority}
\label{Sample-Cam-Image}
\end{center}
\vspace*{-8mm}
\end{figure}

Images can be flexibly retrieved for any desired time and location, and are typically available as a paired set (spanning both directions of traffic flow). An example of the images in this data repository is presented in Figure \ref{Sample-Cam-Image}. 

For this proof-of-concept work, the primary goal is to develop a fully-automated image processing pipeline and assess the feasibility of using the combination of traffic images (exemplified by Singapore's open source data) and this analysis pipeline as a means of quantifying traffic emissions.

\begin{figure}[htbp]
\begin{center}
\centerline{\includegraphics[width=0.9\linewidth]{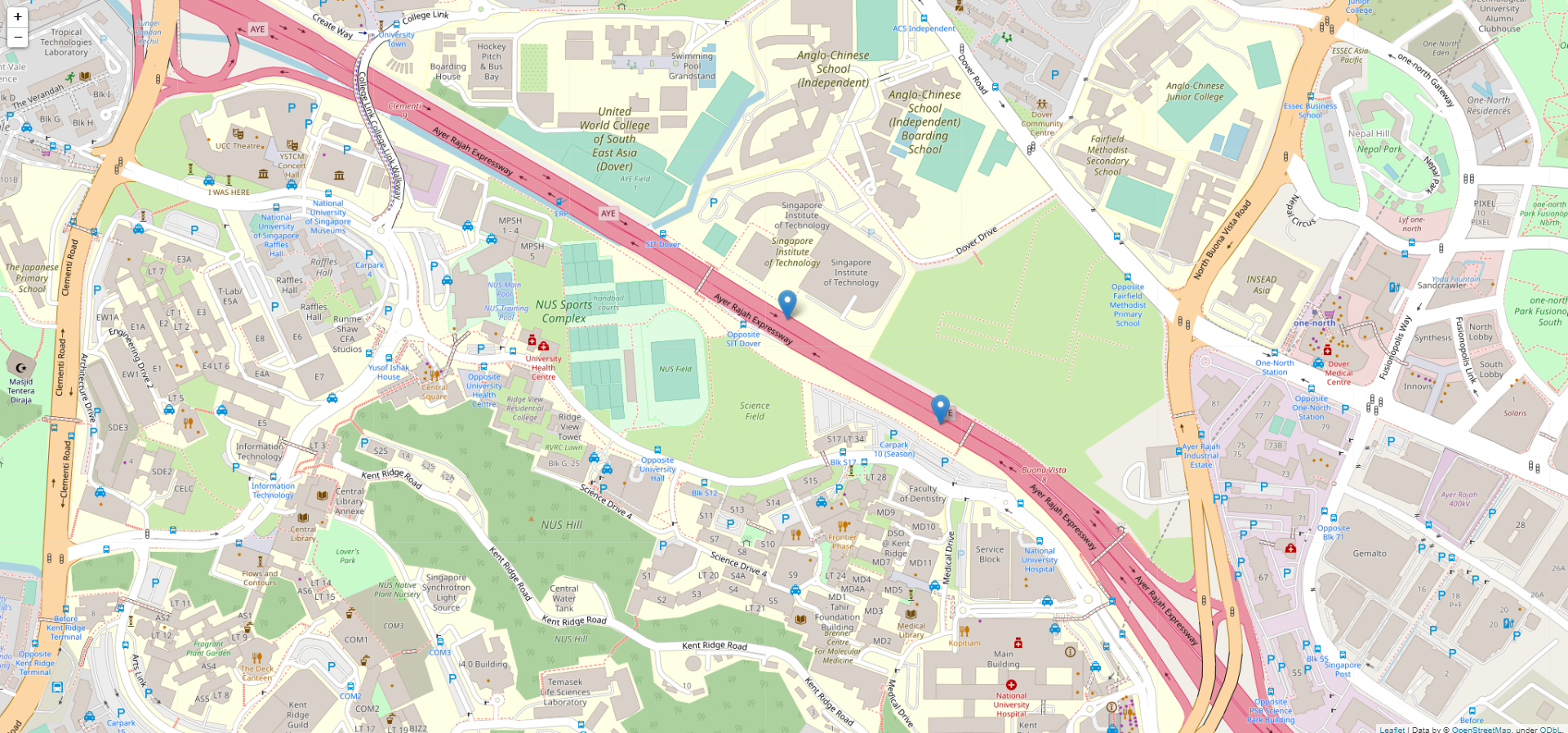}}
\caption{Location of traffic cameras analyzed in this work (blue pins)}
\label{Camera 2}
\end{center}
\vspace*{-7mm}
\end{figure}

Hence, this work focuses on one location, a stretch of the Ayer Rajah Expressway (a high-speed freeway) along the perimeter of the National University of Singapore. The exact location of the traffic cameras identified and analyzed in this work are marked on Google Maps as blue-colored pins in Figure \ref{Camera 2}. Two locations are marked as they provide vehicular counts for traffic flow away from and towards the city centre respectively.

While the repository documentation specifies that images are available every minute, it was discovered during initial data exploration that the traffic cameras do not actually acquire images with a consistent periodicity. Even when a 5 minute interval is used, there is still a significant amount of image repetition (e.g. 40 duplicates in 327 retrieved images, $\approx$ 12\%). More frequent image retrieval from the data repository will consequently result in excessive analyzed images being duplicates. Hence, this limitation in number of images that can be extracted from the traffic cameras constrains the estimate of vehicular counts to time intervals on the order of 5 minutes or greater in this work.

\subsection{Particulate Matter (PM) Measurement at Location of Interest}

Pollutant emission is frequently quantified for various sizes, spanning PM$_{1}$, PM$_{2.5}$ and PM$_{10}$, as they differ substantially in proportion across emission sources and in terms of their physiological effect. Hence, sensors have been set-up along this stretch of road to quantify the PM$_{1}$ and PM$_{2.5}$ concentration levels in addition to Relative Humidity (RH) and Temperature. The sensor acquires data every second (1 Hz acquisition frequency), and multiple sensors can be deployed at different distances from the road simultaneously to acquire concurrent measurements. 

\begin{figure}[htbp]
\begin{center}
\centerline{\includegraphics[width=0.9\linewidth]{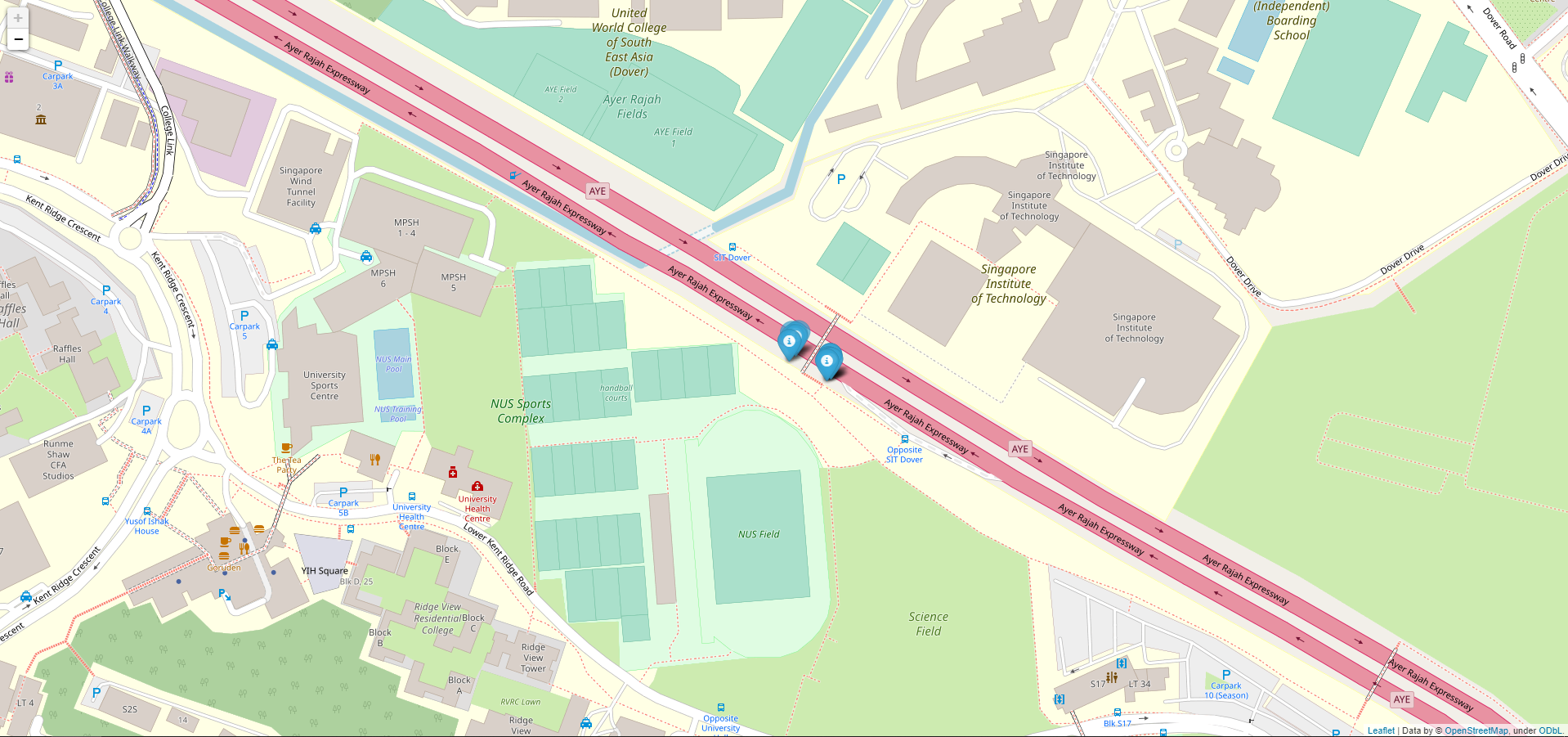}}
\caption{Acquisition location of PM readings in set L1 corresponding to roadside measurements. Each blue pin corresponds to the sensor position for a particular day.}
\label{Pollution-L1}
\end{center}
\vspace*{-8mm}
\end{figure}

\begin{figure}[htbp]
\begin{center}
\centerline{\includegraphics[width=0.9\linewidth]{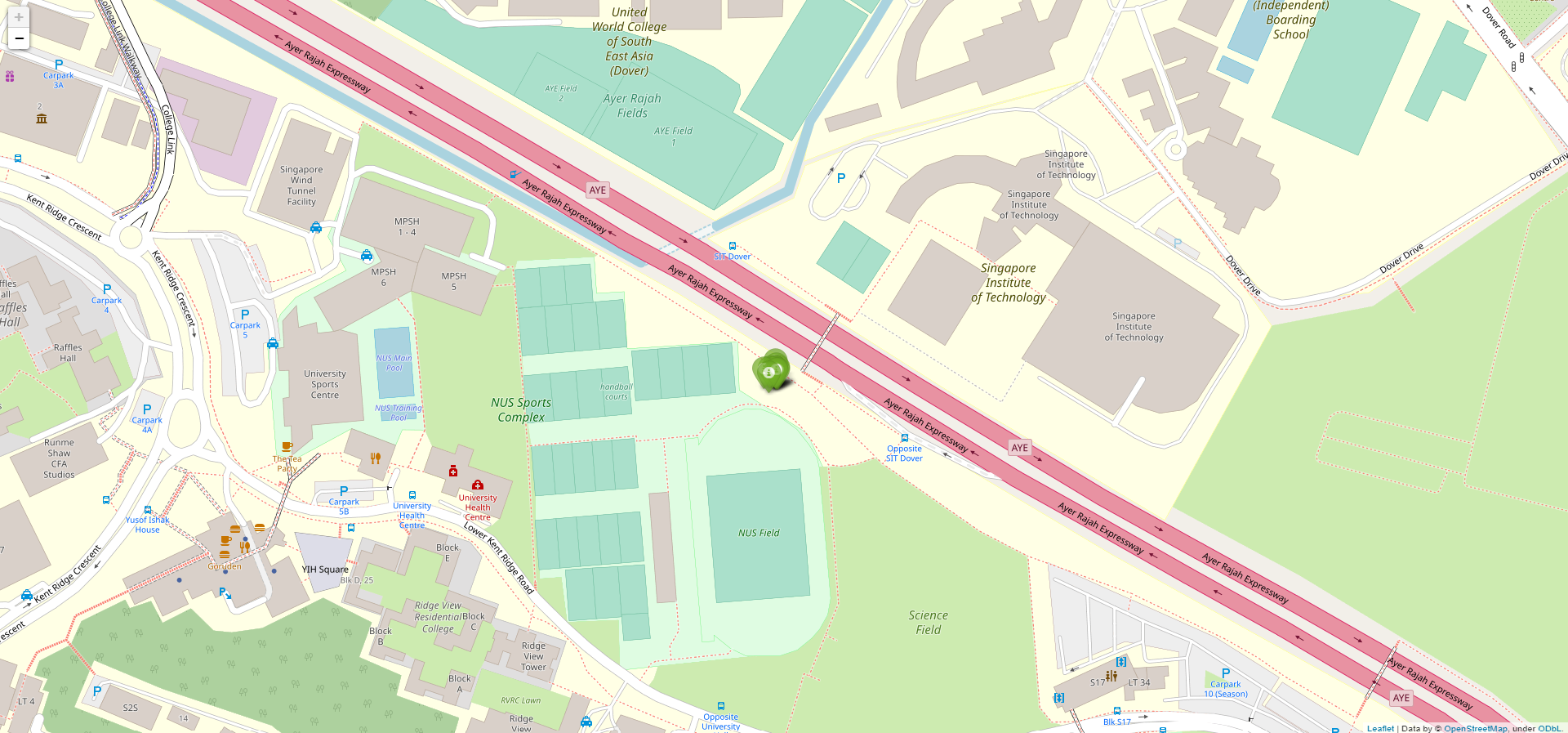}}
\caption{Acquisition location of PM readings acquired in set L2 corresponding to measurements approximately 22m from the roadside. Each green pin corresponds to the sensor position for a particular day.}
\label{Pollution-L3}
\end{center}
\end{figure}

In this work, readings from 2 sensors are analyzed. The first sensor is situated along the roadside (L1), while the second sensor is situated at a distance approximately 22 meters from the road to obtain an estimate of the prevailing baseline pollutant level (L2). The paired set of 2 sensor locations across different days are indicated in Figure \ref{Pollution-L1} (L1) and Figure \ref{Pollution-L3} (L2) respectively. It should be noted that the exact positions are not identical from day to day due to experimental variability despite best efforts to ensure consistency throughout the period of measurement.

Measurements were obtained for similar time periods in the afternoon (typically between 1 - 3 pm) spanning multiple days in February (21 - 25 Feb) and March (4, 5, 7 - 9, 11 Mar) 2022. Although measurements were obtained for 11 days in total, issues with sensor data acquisition across different locations meant that some days had limited amounts of data, and were subsequently excluded from analysis.

In addition, to facilitate consistency of comparison with the vehicular counts obtained from the traffic images derived above, the PM readings were further averaged across 5 minute time blocks despite the 1 Hz acquisition frequency.

\subsection{Traffic Count Pipeline}

An automated pipeline for evaluation of traffic counts is developed in this work and described below. 

Firstly, the images to be utilized for traffic counts are downloaded from the data repository described above \cite{traffic_images}. These images are then masked via an automated Python script to retain only the road section of interest, thereby reducing any extraneous interference during object segmentation and classification. This was found to be helpful for improvement of the vehicle detection and classification algorithm. 

Next, the processed images are fed into a pre-trained object detection algorithm -- Faster RCNN Inception Resnet-v2 (1024x1024) \cite{birodkar_2021}. This model was previously trained on the COCO 2017 dataset with a mAP of 38.7 \cite{lin2014microsoft}. 

\begin{figure}[htbp]
\begin{center}
\centerline{\includegraphics[width=0.9\linewidth]{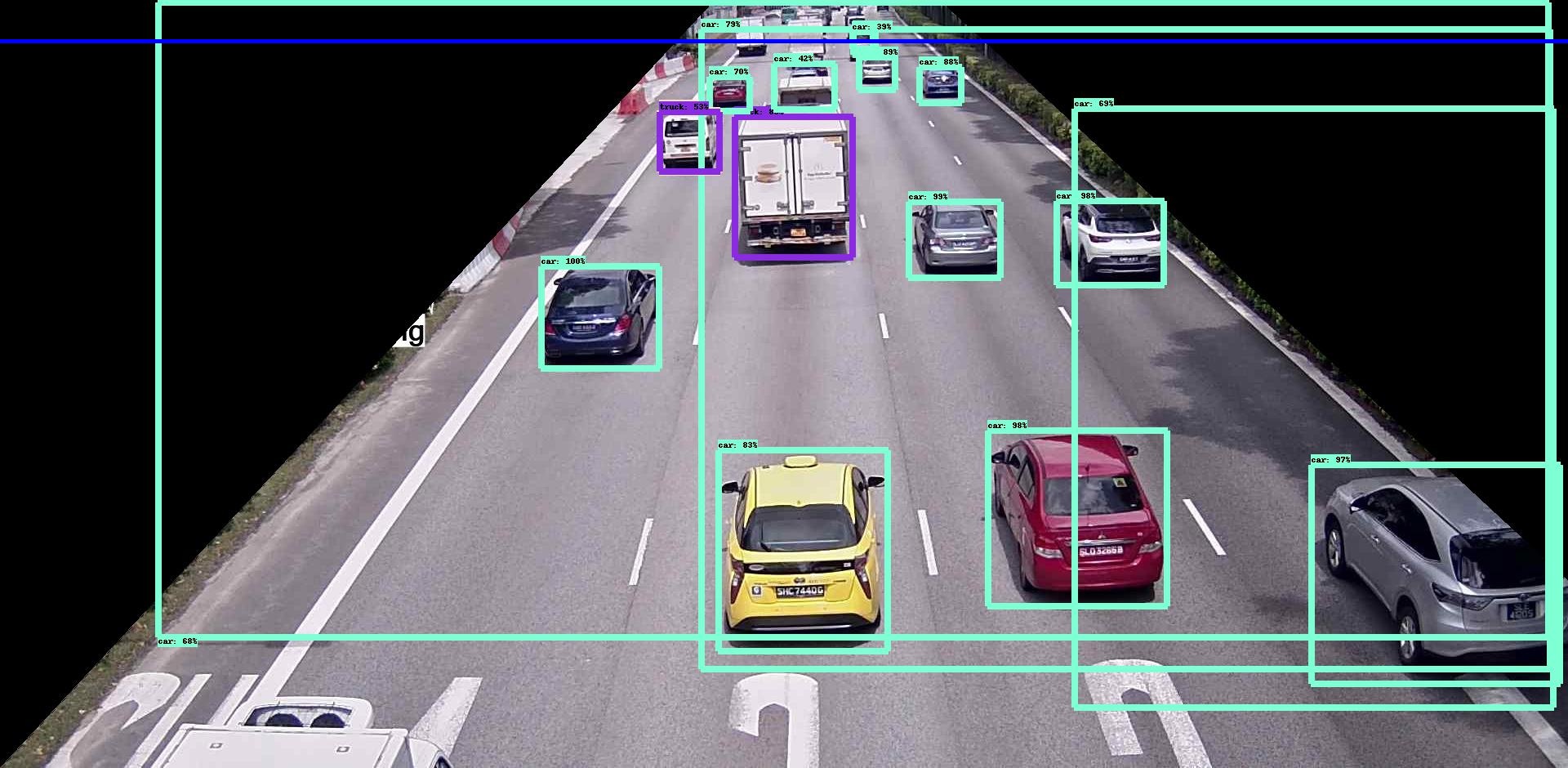}}
\caption{Example of output from the baseline model with erroneous detection boxes}
\label{Example 1}
\end{center}
\vspace*{-8mm}
\end{figure}

\begin{figure}[htbp]
\begin{center}
\centerline{\includegraphics[width=0.9\linewidth]{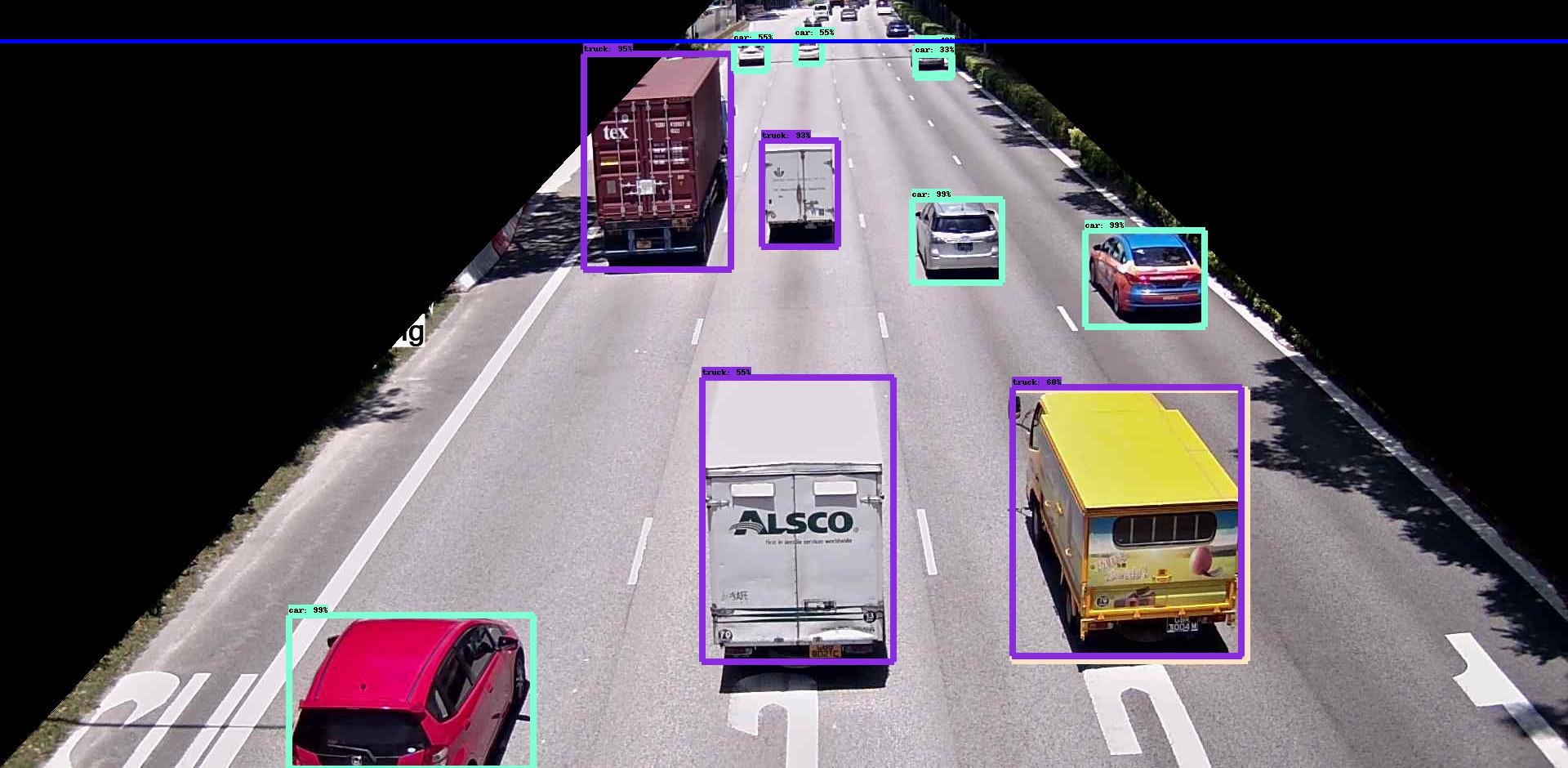}}
\caption{Example of objects detected by the baseline model with multiple classification classes}
\label{Example 2}
\end{center}
\vspace*{-7mm}
\end{figure}

However, the baseline model as-trained exhibits certain accuracy issues, namely double-counting of vehicles (classification into multiple classes) and erroneous detection boxes which are frequently abnormally-sized. Figure \ref{Example 1}, and Figure \ref{Example 2} are illustrative examples of objects identified from the base model with some of these errors.

Hence, further filter functions are applied to enhance model accuracy. These filtering algorithms are implemented on the object segmentation result to remove instances where objects are classified as belonging to multiple classes and instances where the object dimensions are too large or too small. For instances with multiple possible classes, the highest probability class is reported. 

\begin{figure}[htbp]
\begin{center}
\centerline{\includegraphics[width=0.9\linewidth]{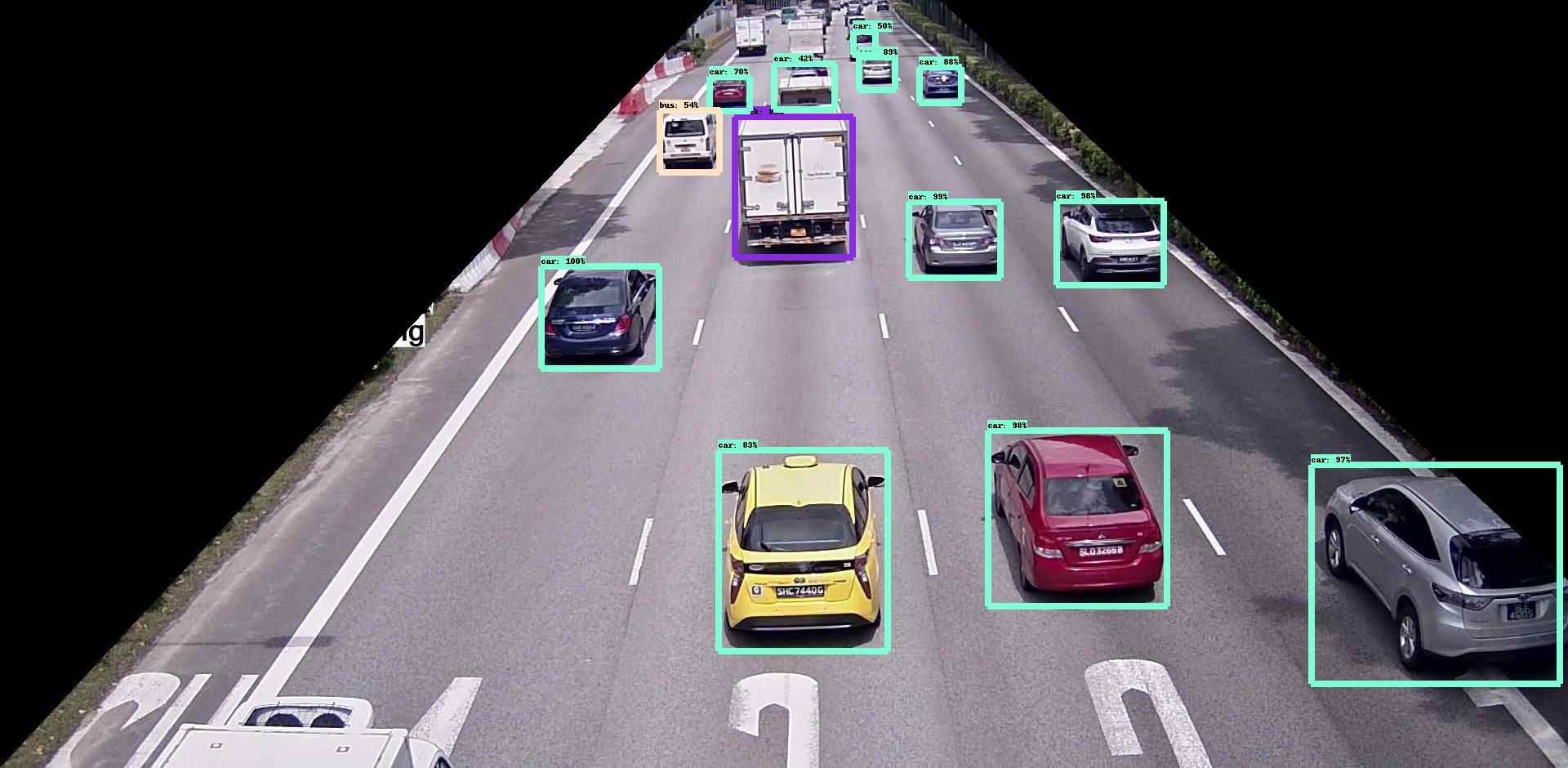}}
\caption{Objects detected after filtering for abnormally-sized objects}
\label{Filter 1}
\end{center}
\vspace*{-8mm}
\end{figure}

\begin{figure}[htbp]
\begin{center}
\centerline{\includegraphics[width=0.9\linewidth]{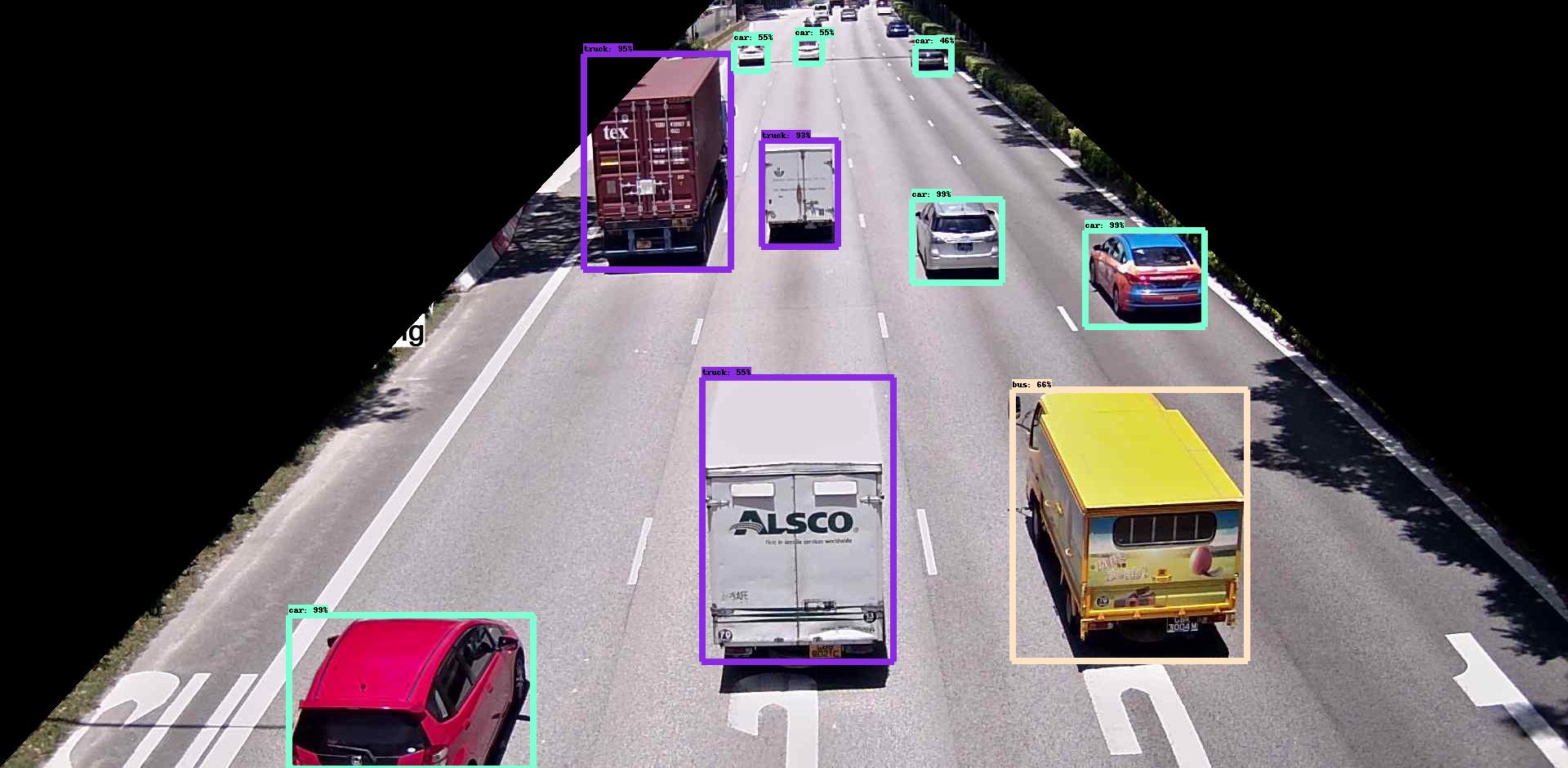}}
\caption{Objects detected after filtering for objects with multiple possible classifications}
\label{Filter 2}
\end{center}
\vspace*{-6mm}
\end{figure}

Figure \ref{Filter 1} and Figure \ref{Filter 2} illustrate the reduction in such instances after applying these filter functions.

Critically, we note that the current model inherently allows for identification and classification of motorcycles, cars, trucks and buses, which span the major categories of vehicles on the road. However, for the purposes of this work, motorcycles are not identified and analyzed as vehicular emissions is anticipated to be dominated by larger vehicles such as cars, trucks and buses. In addition, we note that Singapore's vehicular population is primarily dominated by sedan cars, as is apparent from the sample images above.

\section{Results}

As described in Methods, a time interval of 5 minutes is used for the aggregation of traffic counts and PM levels in the following analysis. While the PM readings are acquired at much higher frequencies (1 Hz), the sensor values are averaged across 5 minutes due to limitations in the availability of traffic images for analysis.

\subsection{Performance of Vehicle Count Algorithm}

As part of model validation, a set of 15 random images were downloaded and manually labeled by the authors. These then served as the ground truth for validating the performance of the automated image analysis pipeline for vehicular counts.

From these set of images, we note that 75\% of the vehicles present in the images were cars, with the other 20\% being trucks and 5\% being buses. Hence, the ability to accurately identify cars in the images is disproportionately important as they are the dominant vehicle on the road.

The detection errors for the baseline model is tabulated for this set of images in Table \ref{tab: Model-Error}. In general, the baseline model is able to detect almost all the cars, trucks and buses, being able to correctly identify more than 90\% of the cars in the images. However, it does miss approximately 4\% of the cars. 

In addition, we note that the baseline model may over-estimate the vehicular counts by between 10 and 23\% as it tends to mis-classify various objects in the images as potential cars and/or other vehicles. 

\begin{table}[htbp]
\caption{Vehicle Count Detection Errors for Baseline Model}
\begin{center}
\begin{tabular}{|c|c|c|c|}
\hline
{Vehicle} & {Correctly} & {Undetected} & {Falsely} \\
{Class} & {Identified} & & {Detected} \\
\hline
{Car} & {$0.941$} & {$0.039$} & {$0.105$} \\
\hline
{Trucks $\&$ Buses} & {$0.808$} & {$0.154$} & {$0.226$} \\
\hline
\end{tabular}
\label{tab: Model-Error}
\end{center}
\end{table}

However, after implementation of the additional filtering algorithms described above, the errors in false detection can be greatly reduced, albeit at the cost of a slight drop in the number of correctly identified vehicles. The collated results in Table \ref{tab: Model-Error-Update} show that falsely detected vehicles are reduced by between 2-3x relative to before. This thus ensures that the vehicular count estimates obtained through this pipeline are more representative of the underlying ground truth.

\begin{table}[htbp]
\caption{Vehicle Count Detection Errors with Additional Filters}
\begin{center}
\begin{tabular}{|c|c|c|c|}
\hline
{Vehicle} & {Correctly} & {Undetected} & {Falsely} \\
{Class} & {Identified} & & {Detected} \\
\hline
{Car} & {$0.895$} & {$0.078$} & {$0.052$} \\
\hline
{Trucks $\&$ Buses} & {$0.769$} & {$0.154$} & {$0.077$} \\
\hline
\end{tabular}
\label{tab: Model-Error-Update}
\end{center}
\end{table}

\subsection{PM Levels and Vehicle Counts for each day}

\begin{figure}[htbp]
\begin{center}
\centerline{\includegraphics[width=0.9\linewidth]{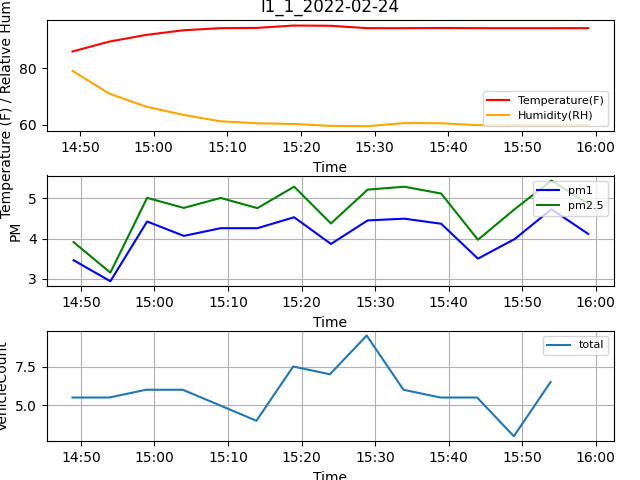}}
\caption{Plots of relative humidity, temperature, PM$_{1}$, PM$_{2.5}$ and vehicular counts recorded.}
\label{242}
\end{center}
\end{figure}

After verification of the accuracy of the automated vehicular detection model, the entire pipeline is used to provide concurrent vehicular counts for the days and times corresponding to the PM data acquisition experiments. Sample graphs are presented for the data acquired on 24 Feb 2022. Figure \ref{242} illustrates traffic counts for the same time period as the sensor's PM, RH and Temperature measurements. 

\begin{figure}[htbp]
\begin{center}
\centerline{\includegraphics[width=0.9\linewidth]{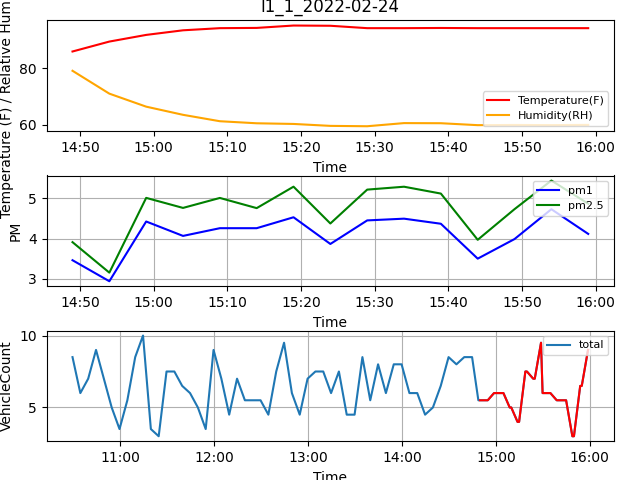}}
\caption{Plots of relative humidity, temperature, PM$_{1}$, PM$_{2.5}$ and vehicular counts recorded. The red solid line indicates the exact time for which the vehicular count corresponds to the PM sensor data.}
\label{242-2}
\end{center}
\vspace*{-7mm}
\end{figure}

Further, Figure \ref{242-2} illustrates the results for traffic counts starting from earlier in the day, prior to the start of sensor data acquisition. This second graph further highlights the fairly stable vehicular counts prior to the start of PM data acquisition, alleviating concerns that concurrent traffic density analysis may be misleading as traffic density is potentially a leading variable relative to particulate emission levels. 

In addition, it is worth noting that the ease of vehicular count data acquisition here further highlights the utility of this data analysis pipeline as a simple and effective means of quantifying vehicular count and potentially, traffic-related particulate emissions.

From the prior graphs, we note that the PM$_{1}$ and PM$_{2.5}$ values remain fairly consistent over the selected period of measurement. In addition, the vehicular counts obtained across each day are also very consistent. This is unsurprising as one would expect the traffic density across each weekday afternoon to not differ substantially. 

\begin{figure}[htbp]
\begin{center}
\centerline{\includegraphics[width=0.9\linewidth]{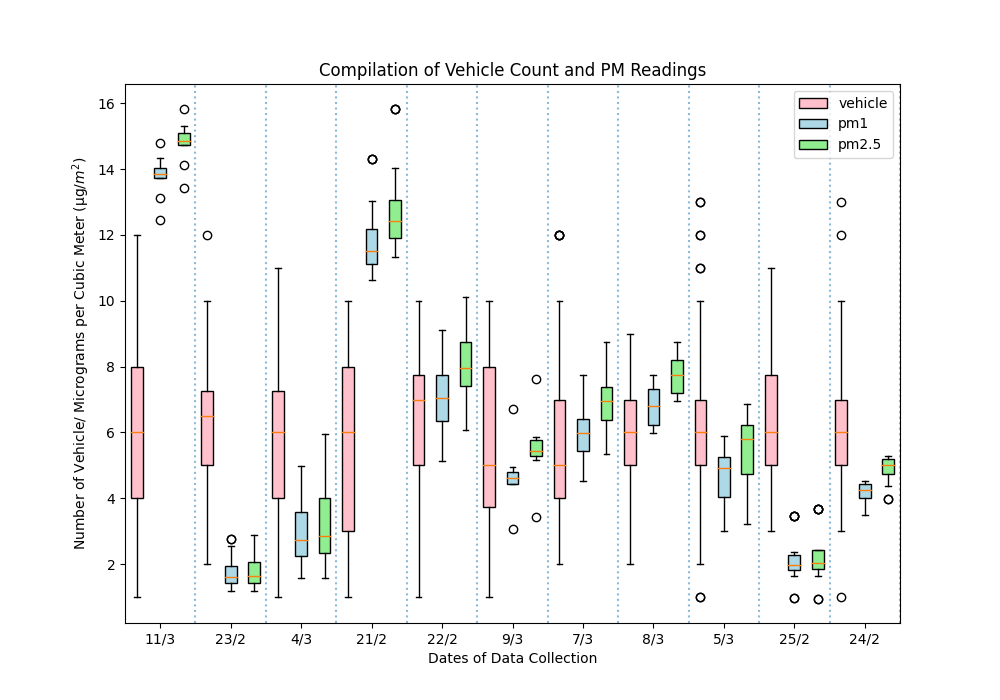}}
\caption{Boxplot of daily traffic counts and PM readings}
\label{boxplot 1}
\end{center}
\vspace*{-8mm}
\end{figure}

Box plots of traffic counts and PM readings across the different days for location L1 are further plotted in Figure \ref{boxplot 1} to illustrate the variability in particulate matter from day to day, and the fairly consistent variation in particulate matter concentrations for both PM$_{1}$ and PM$_{2.5}$.

\begin{figure}[htbp]
\begin{center}
\centerline{\includegraphics[width=0.9\linewidth]{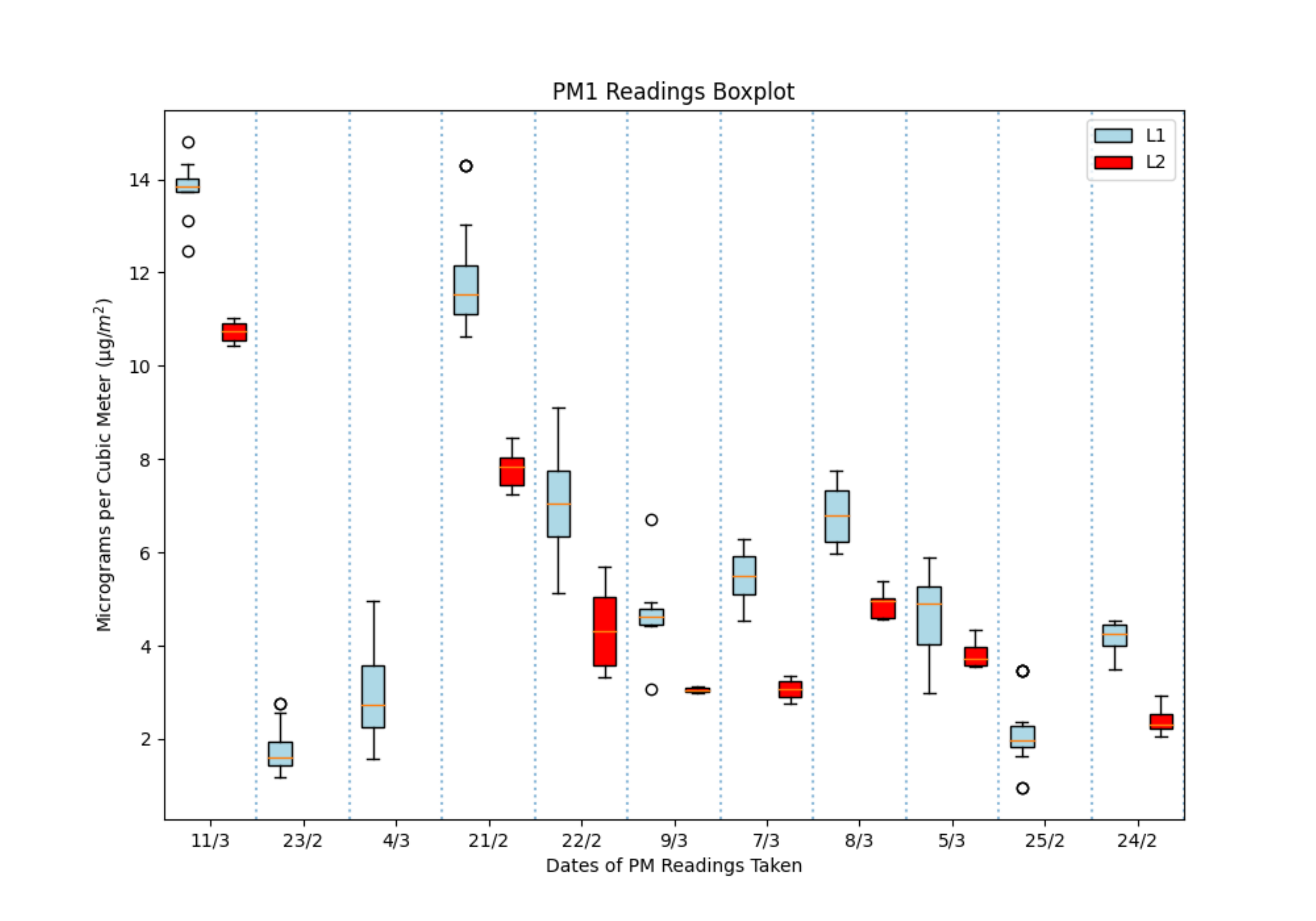}}
\caption{Boxplot of PM$_{1}$ levels (L1) next to and (L2) away from the road}
\label{boxplot-Locations}
\end{center}
\vspace*{-8mm}
\end{figure}

For ease of analysis, and capitalizing on the fairly consistent levels in PM$_{1}$ and PM$_{2.5}$ observed above, we restrict the subsequent analysis to the PM$_{1}$ values. Box plots of the PM$_{1}$ readings across the different days for both location L1 and L2 are plotted in Figure \ref{boxplot-Locations} to illustrate the variation in particulate matter at the road and away from the road. It is apparent that the PM$_{1}$ values are elevated at the road due to the vehicular emissions, as is consistent with prior reports from literature. In addition, due to experimental difficulties, we note that not all days have data for both L1 and L2.

\subsection{Correlation Between Vehicular Counts and PM Levels}

\begin{figure}[htbp]
\begin{center}
\centerline{\includegraphics[width=0.9\linewidth]{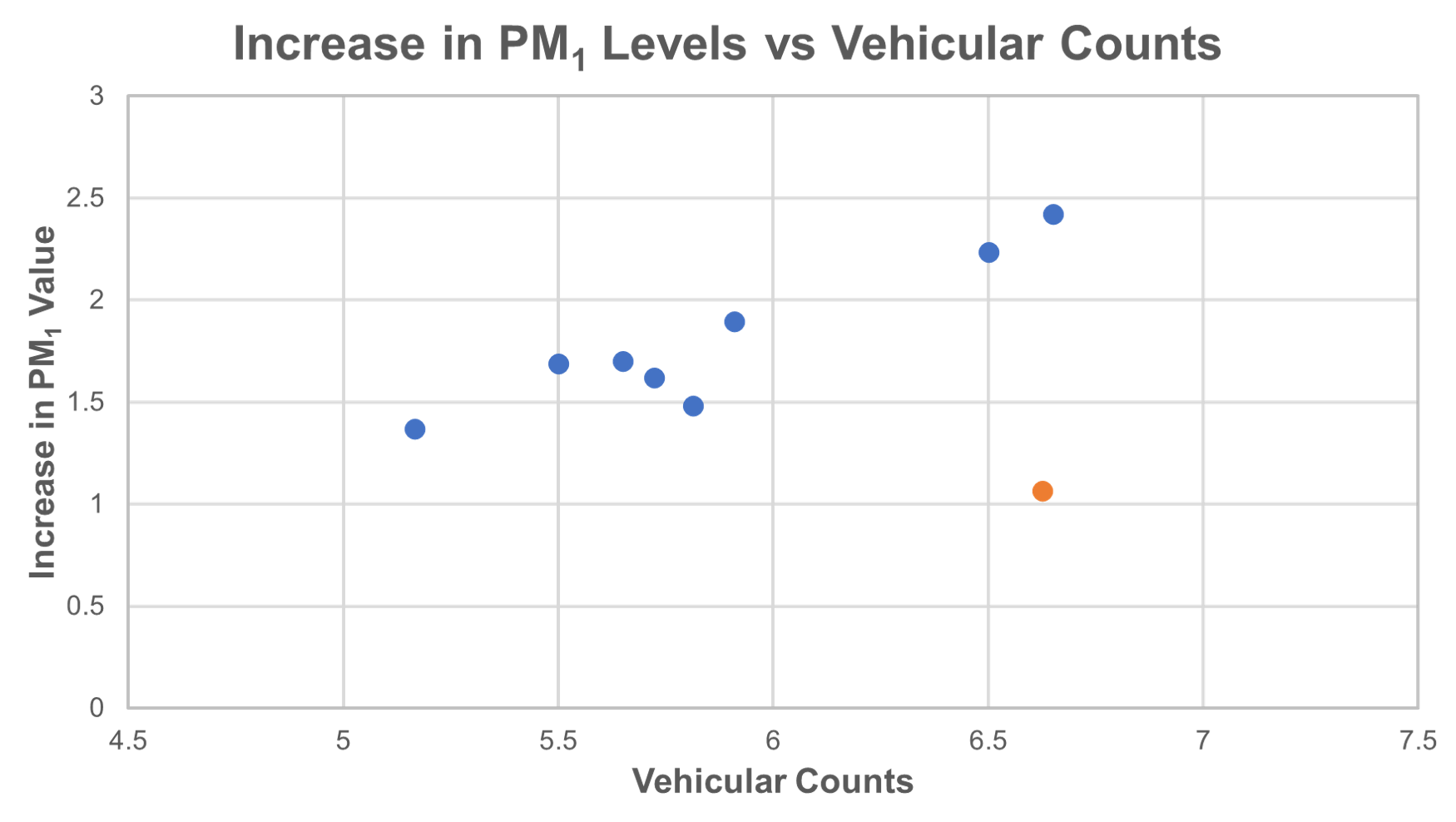}}
\caption{Plot of Increase in PM$_{1}$ levels vs Vehicular Counts}
\label{correl-plot}
\end{center}
\end{figure}

Using the PM$_{1}$ values measured at L2 as the prevailing baseline for each day, we calculate the difference in measured PM$_{1}$ value across locations L1 and L2 for each day as a proxy for the approximate increase in particulate matter due to vehicular emissions. We further calculate the correlation between the increase in particulate matter and the average vehicular count on each day.The increase in PM$_{1}$ value and the vehicular count for each day is plotted in Figure \ref{correl-plot}. Excitingly, we note that a Pearson's correlation coefficient of 0.93 is obtained when a single outlier (23rd Feb) is removed from the analysis. 

This thus confirms that the use of this automated image processing pipeline as a proxy for vehicular pollutant emission is returning physically relevant and accurate estimates.

\section{Conclusion}

In this work, we developed an automated pipeline to analyze traffic camera-derived images along a major expressway in Singapore to obtain vehicular counts. This pipeline rapidly counts the number of cars and heavy vehicles (e.g. buses and trucks) in the image, and provide estimates of vehicle density along specific road segments. This can be useful for purposes such as monitoring the pollutant emission from traffic in order to assess the efficacy of various intervention measures such as new traffic regulations. 

Our results indicate good correlation of $>$ 0.9 between the vehicular counts and the increase in particulate matter levels at the road relative to prevailing baseline levels far from the road, further indicating the utility of such methods for automated pollutant monitoring. Nonetheless, it should be noted that vehicle count is an indirect estimate, and is contingent on approximations regarding the level of emissions per vehicle type at specific traveling velocities. Hence, complementary actual measurements of emissions may still be needed in certain instances as a form of calibration.

In addition, we note that there are not insignificant differences in the locations of data collection for traffic count and PM readings, which may potentially account for some deviation in the PM readings relative to the vehicle count. Better collocation of the PM sensors and camera might further improve the correlation described in this work. Warm and humid conditions such as those prevalent in Singapore can also cause some sensor deviation. More frequent sensor calibrations during the period of data acquisition may also improve this comparison.

While promising, this remains a proof-of-concept, and we note certain deficiencies in the model pipeline developed. Firstly, this automated pipeline does not take into account any deficiencies in the image data-set. For example, images in this particular open-source repository are only available at $\approx 5$ minute intervals and this may be excessively large for some purposes. More frequent image acquisition may provide better estimates of the instantaneous vehicular density and capture finer variations in the pollutant emission. While only one data-set source was tested, we believe this pipeline to be highly flexible to various sources of image data.

Lastly, we note that the neural network model used can be improved further if more resources are available to acquire additional labeled data to further train and optimize the model. In particular, the ability to differentiate taxis and/or motorcycles can be helpful for analysis. Nonetheless, we believe the results here to be very promising in demonstrating the use of this machine learning pipeline as a means of quantifying vehicles on the road, and consequently, the additional pollution that may result.

\bibliography{mybibfile}

\begin{thebibliography}{10}
\expandafter\ifx\csname url\endcsname\relax
  \def\url#1{\texttt{#1}}\fi
\expandafter\ifx\csname urlprefix\endcsname\relax\def\urlprefix{URL }\fi
\expandafter\ifx\csname href\endcsname\relax
  \def\href#1#2{#2} \def\path#1{#1}\fi

\bibitem{buckeridge2002effect}
D.~L. Buckeridge, R.~Glazier, B.~J. Harvey, M.~Escobar, C.~Amrhein, J.~Frank,
  Effect of motor vehicle emissions on respiratory health in an urban area.,
  Environmental health perspectives 110~(3) (2002) 293--300.

\bibitem{quah2003economic}
E.~Quah, T.~L. Boon, The economic cost of particulate air pollution on health
  in singapore, Journal of Asian Economics 14~(1) (2003) 73--90.

\bibitem{nea|air_quality}
\href{https://www.nea.gov.sg/our-services/pollution-control/air-pollution/air-quality}{Air
  quality}.
\newline\urlprefix\url{https://www.nea.gov.sg/our-services/pollution-control/air-pollution/air-quality}

\bibitem{shewmake2012can}
S.~Shewmake, Can carpooling clear the road and clean the air? evidence from the
  literature on the impact of hov lanes on vmt and air pollution, Journal of
  Planning Literature 27~(4) (2012) 363--374.

\bibitem{daniel2000environmental}
J.~I. Daniel, K.~Bekka, The environmental impact of highway congestion pricing,
  Journal of Urban Economics 47~(2) (2000) 180--215.

\bibitem{cai2007estimation}
H.~Cai, S.~Xie, Estimation of vehicular emission inventories in china from 1980
  to 2005, Atmospheric Environment 41~(39) (2007) 8963--8979.

\bibitem{davis2005development}
N.~Davis, J.~Lents, M.~Osses, N.~Nikkila, M.~Barth, Development and application
  of an international vehicle emissions model, Transportation Research Record
  1939~(1) (2005) 156--165.

\bibitem{pokharel2002road}
S.~S. Pokharel, G.~A. Bishop, D.~H. Stedman, An on-road motor vehicle emissions
  inventory for denver: an efficient alternative to modeling, Atmospheric
  Environment 36~(33) (2002) 5177--5184.

\bibitem{boppana2019cfd}
V.~B. Boppana, D.~J. Wise, C.~C. Ooi, E.~Zhmayev, H.~J. Poh, Cfd assessment on
  particulate matter filters performance in urban areas, Sustainable Cities and
  Society 46 (2019) 101376.

\bibitem{konczak2021assessment}
B.~Ko{\'n}czak, M.~Cempa, M.~Deska, et~al., Assessment of the ability of
  roadside vegetation to remove particulate matter from the urban air,
  Environmental Pollution 268 (2021) 115465.

\bibitem{velasco2016particles}
E.~Velasco, S.~H. Tan, Particles exposure while sitting at bus stops of hot and
  humid singapore, Atmospheric Environment 142 (2016) 251--263.

\bibitem{zheng2021impacts}
T.~Zheng, H.-W. Wang, X.-B. Li, Z.-R. Peng, H.-D. He, Impacts of traffic on
  roadside particle variations in varied temporal scales, Atmospheric
  Environment 253 (2021) 118354.

\bibitem{traffic_images}
\href{https://data.gov.sg/dataset/traffic-images}{Traffic images-data.gov.sg}.
\newline\urlprefix\url{https://data.gov.sg/dataset/traffic-images}

\bibitem{birodkar_2021}
\href{https://github.com/tensorflow/models/blob/master/research/object_detection/g3doc/tf2_detection_zoo.md}{Tensorflow
  2 detection model zoo} (May 2021).
\newline\urlprefix\url{https://github.com/tensorflow/models/blob/master/research/object_detection/g3doc/tf2_detection_zoo.md}

\bibitem{lin2014microsoft}
T.-Y. Lin, M.~Maire, S.~Belongie, J.~Hays, P.~Perona, D.~Ramanan,
  P.~Doll{\'a}r, C.~L. Zitnick, Microsoft coco: Common objects in context, in:
  European conference on computer vision, Springer, 2014, pp. 740--755.

\end{thebibliography}

\end{document}